\documentclass[proceedings, preprint]{rmaa}

\usepackage{paralist}
\newcommand{\D}{\discretionary{}{}{}}

\usepackage{psfrag,color}
\usepackage{dsfont}
\usepackage{amsmath,amsfonts,amsthm,bm}

\SetYear{2022}
\SetConfTitle{Prospects for low-frequency radio astronomy in S. A.}

\title{A sparse coding approach to inverse problems with application to microwave tomography}

\author{
  C. F. Caiafa\altaffilmark{1} and
  R. M. Irastorza\altaffilmark{2}}

\altaffiltext{1}{Instituto Argentino de Radioastronom\'i{}a,
  CONICET CCT La Plata/CIC-PBA/UNLP, V. Elisa, Argentina
  (ccaiafa\D{}@fi.\D{}uba.\D{}ar).}
\altaffiltext{2}{Instituto de F\'i{}sica de L\'i{}quidos y Sistemas Biol\'o{}gicos,
  CONICET CCT La Plata/UNLP, La Plata, Argentina
  (rirastorza\D{}@frlp.\D{}utn.\D{}edu.\D{}ar).}

\shortauthor{Caiafa \& Irastorza}
\shorttitle{Sparse Coding for Inverse Problems}

\listofauthors{C. F. Caiafa \& R. M. Irastorza}
\abstract{Inverse imaging problems that are ill-posed can be encountered across multiple domains of science and technology, ranging from medical diagnosis to astronomical studies. To reconstruct images from incomplete and distorted data, it is necessary to create algorithms that can take into account both, the physical mechanisms responsible for generating these measurements and the intrinsic characteristics of the images being analyzed.  In this work, the sparse representation of images is reviewed, which is a realistic, compact and effective generative model for natural images inspired by the visual system of mammals. It enables us to address ill-posed linear inverse problems by training the model on a vast collection of images. Moreover, we extend the application of sparse coding to solve the non-linear and ill-posed problem in microwave tomography imaging, which could lead to a significant improvement of the state-of-the-arts algorithms.} %Microwave tomography can potentially provide a valuable, low-cost and non-invasive imaging tool with applications in medicine and other industrial fields. 
\resumen{En diversas áreas científicas y tecnológicas, desde el diagnóstico de enfermedades por imágenes a estudios de astronomía, nos podemos encontrar con problemas inversos ``mal planteados''. Para reconstruir imágenes a partir de datos incompletos o distorsionados, es necesario que los algoritmos tengan en cuenta el mecanismo físico por el cual se toman las mediciones como así también las propiedades intrínsecas de las imágenes. En este trabajo, se revisa la representación ``rala'' de imágenes como un modelo generativo realista, compacto y efectivo inspirado en el sistema visual de los mamíferos. Este modelo permite resolver el problema lineal inverso ``mal planteado'' entrenándolo sobre un conjunto grande de imágenes. Además, se aplica este modelo para resolver el problema inverso no-lineal y ``mal planteado'' de la tomografía de microondas que podría producir mejoras significativas al estado del arte de esta tecnología.} %La tomografía por microondas podría proveer una herramienta valiosa, no-invasiva y de bajo costo con aplicaciones en la medicina y otras industrias. 
\addkeyword{methods: numerical}
\addkeyword{techniques: image processing}

\begin{document}
\setlength{\abovedisplayskip}{5pt}
\setlength{\belowdisplayskip}{5pt}
% Typeset article header
\maketitle

\section{Introduction}\label{sec:intro}
% Define inverse problem in imaging with examples of interferometry in radiostronomy, MRI, X-ray Tomography and mention Microwave Tomography that will be introduced later.
Solving an inverse problem means to infer the input of a system given its output. In imaging, it refers to obtaining an image of an object or scene from data collected by a device such as an x-ray machine, MRI scanner or array of antennas. While in general modelling the ``direct'' problem, i.e. going from the object to the image, is well known and easy to solve numerically, its ``inverse'' counterpart can be challenging since the measurements are often incomplete, noisy, or indirect, and the underlying image can have infinite solutions that are consistent with the measurements. Solving inverse problems in imaging has many practical applications in medicine, astronomy, geology, and other fields.

\begin{figure}[!t]
\centering
  \includegraphics[width=0.75\columnwidth]{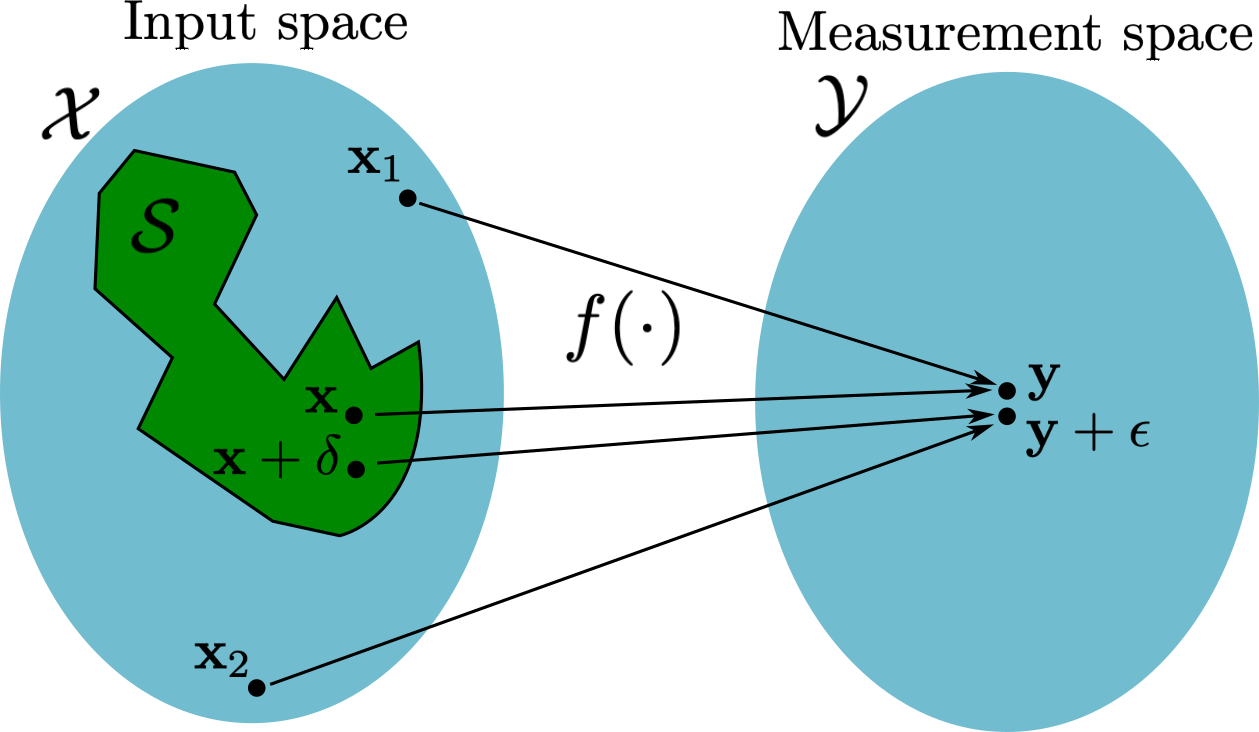}
  \caption{\footnotesize Ill-posed inverse problem: Given a measurement $\mathbf{y}\in \mathcal{Y}$, there are multiple solutions: $\mathbf{x}$ and $\mathbf{x}_1 \in \mathcal{X}$ s.t. $\mathbf{y}=f(\mathbf{x})=f(\mathbf{x}_1)$ (non-uniqueness). Small errors in the measurements may produce large errors in the reconstructions, i.e. $\mathbf{y}+\mathbf{\epsilon} = f(\mathbf{x}_2)$ with large $\|\mathbf{x}_2 - \mathbf{x}\|$ (instability). To avoid these problems we need to restrict the solutions to an apropriate subset $\mathcal{S} \subset \mathcal{X}$.} 
  \label{fig:Fig_1}
\end{figure}

% Define ill-posedness according to Hadamard (1923).
Mathematically, the output $\mathbf{y}$ of a system (measurements) is obtained by the application of some known operator $f(\cdot)$ on the input image $\mathbf{x}$, i.e. 
\begin{equation}\label{eq:direct_problem}
    \mathbf{y} = f(\mathbf{x}).
\end{equation}
% Discussion about How to convert ill-posed problem to a well-posed one? Make a figure showing how to reduce the set of acceptable solutions in order to avoid non-uniqueness.
% - Classical (Tikhonov, total variation)
% - New trend (learn from large collections of examples): deep CNN, GANs.
Inverse problems in imaging are often ill-posed, which means that: 1) there is not a unique solution $\mathbf{x}$ for a given set of measurements $\mathbf{y}$; and 2) the solution does not depend continuously on the given measurements, so noise in the data can lead to large errors in the solution \citep{Hadamard1902} (Fig. \ref{fig:Fig_1}). 

Methods to convert an ill-posed problem into a well-posed problem usually restrict the input set $\mathcal{X}$ to a subset of useful solutions $\mathcal{S}$ as illustrated in Fig. \ref{fig:Fig_1}. 
% Explain that in a machine learning approach, rather than imposing a theoretical model about the images, it provides algorithms that are able to learn the models from large collections of data samples. In this work we focus on Sparse coding method.
To that end, traditional techniques use the Tikhonov regularizer \citep{Tikhonov1977}, minimize Total Variation - TV \citep{Rudin1992} or use others regularizers that impose theoretical properties on the reconstructed images \citep{Bertero2001}. On the other side, more recent approaches rely on large datasets to train deep neural networks for generating plausible input images as solutions \citep{Unser2017}. In the latter, there is no theoretical model about the images. The neural network is used as a black-box that map output measurements to input images.
% Improve this explanation and highlight the advantages of this approach compared to traditionala and black box NNs.
In this work, we focus on the sparse coding model of images, which can be used to solve ill-posed inverse problems and avoid the issues of the black-box neural networks. 

\section{Sparse coding model of images} \label{sec:sparse_coding}
\begin{figure}[!t]
\centering
  \includegraphics[width=0.80\columnwidth]{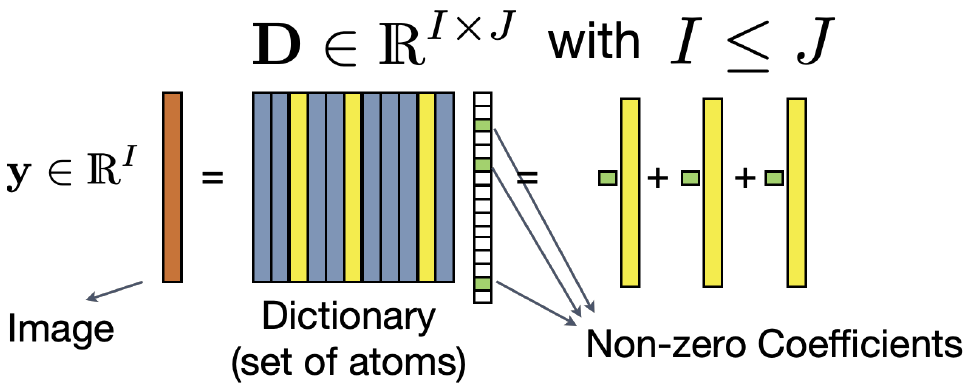}
  \caption{\footnotesize Sparse coding model: every image of interest can be written as a linear superposition of a few image prototypes (atoms) in a dictionary.} 
  \label{fig:Fig_2}
\end{figure}

% Show evidence in biological visual system (Hubel&Wiesel, 1968)
% Introduce mathematics and figures for sparse coding model
% Important questions: 1) How to compute the sparse vector; 2) Is a sparse representation unique?; 3) How to choose good dictionaries.
Sparse coding of images is inspired by the visual system of mammals after the work of \citet{Hubel&Wiesel1979} for which they were awarded with the Nobel prize in 1981. They discovered that specific patterns presented in the visual field activate neurons in the primary visual cortex (V1), resulting in encoding of the image in the brain with only a few neurons in the V1 cortex being activated.
%neurons in the primary visual cortex (V1) are designed to activate when particular patterns are presented in the view field so an image is encoded in the brain as a small number of activated neurons in the V1 cortex. 
In the sparse coding model, if we represent an image having $I$ pixels as a vector $\mathbf{x}\in\mathds{R}^I$, then we assume that it can be written as a linear combination of a few elementary patterns (atoms) selected from a dictionary $\mathbf{D}\in\mathds{R}^{I\times J}$ ($J\ge I$). Mathematically, we write:
\begin{equation}\label{eq:sparse_coding}
    \mathbf{x} = \mathbf{D}\mathbf{s}, \mbox{ with } \|\mathbf{s}\|_0 \le K \ll J,
\end{equation}
where $\mathbf{s}\in\mathds{R}^J$, $J\ge I$ and $\|\mathbf{s}\|_0$ gives the number of non-zero entries in vector $\mathbf{s}$ (Fig. \ref{fig:Fig_2}).

The sparse coding model allows for compression of natural images since they can be represented by a few (small $K$) coefficients when an appropriate dictionary is chosen. When the dictionary $\mathbf{D}$ is learned from a large image dataset, the obtained atoms emerge as a family of localized, oriented, band-pass receptive fields, similar to those found in the primary visual cortex \citep{Olshausen&Field1996}.

During the last years the sparse coding theory was developed and various techniques were proposed to solve different image processing problems. The following theoretical and practical questions emerged and motivated the research in this field called Compressed Sensing (CS) \citep{Foucart2013}:

\noindent {\it (a) Given an image $\mathbf{x}$ and a chosen dictionary $\mathbf{D}$, how can we compute the sparse vector of coefficients $\mathbf{s}$ such that eq. (\ref{eq:sparse_coding}) is satisfied?}
A wide range of algorithms are available now for this purpose. Some of them proceed by looking for each of the non-zero coefficients, one by one, using a deflation technique \citep{Tropp2008}. Other methods use an $\ell_1$-norm regularization term which has the effect of promoting sparse solutions \citep{Donoho2003}. These methods are referred as basis pursuit. Finally, there are a family of algorithms that iteratively apply thresholding of the solutions which also impose sparsity \citep{Beck2009}.
    
    % start form the simple orthogonal dictionary case and  relax it to more general algorithms (MP and BP)
\noindent {\it (b) Is the obtained sparse vector $\mathbf{s}$ unique?}
    % Introduce some sufficient conditions
 When applying sparse representation to restrict the solutions of an ill-posed inverse problem, it is important to know if this constraint allows for uniqueness of solutions. Several sufficient conditions were found, such that, if met then the solution of equation (\ref{eq:sparse_coding}) is unique \citep{Eldar2012}.
    
\noindent {\it (c) How to chose a ``good'' dictionary $\mathbf{D}$ for a given dataset?} Dictionaries specially designed to efficiently compress natural images were developed based on mathematical operators such as the Discrete Cosine Transform (DCT) and Wavelet Transform (WT). Also, dictionary learning algorithms were proposed relying on available large datasets of images, thus achieving higher compression rates (sparser vector of coefficients) \citep{Elad2014}.

\noindent {\it (d) How to extend sparse coding to multidimensional signals?} Finding sparse representations is a computationally intensive task and its complexity scales exponentially with the dimensionality of the signals. A method to break the {\it curse of dimensionality} for higher dimensional signals, usually known as tensors, is to consider a dictionary having a Kronecker structure \citep{Caiafa2013}.
    % Show the curse of dimensionality effect when computing sparse representation of higher order tensors. Introduce sparse tucker model to avoid the curse of dimensionality

\section{Sparse Coding applied to inverse linear imaging problems}\label{sec:linear}
% To present the general approach for any linear system
% Then, to show selected examples with references for different linear problmes
Ill-posed inverse problems can be found even in the simplest mathematical form of equation (\ref{eq:direct_problem}). In this section, we show how the sparse coding model can help finding useful solutions of ill-posed inverse linear problems such as in image completion, superresolution and others.

Let us assume that we want to recover some image $\mathbf{x}\in\mathds{R}^I$ and we have at our disposal a set of $M<I$ linear measurements, i.e. $\mathbf{y}\in\mathds{R}^M$ obtained as follows:
\begin{equation}\label{eq:linear_eq}
    \mathbf{y} = \mathbf{\Phi}\mathbf{x},
\end{equation}
where $\mathbf{\Phi}\in\mathds{R}^{M\times I}$ is some linear operator.

When the number of measurements is smaller than the size of the signal to recover ($M<I$), basic linear algebra tell us that there is an infinite number of vectors $\mathbf{x}$ satisfying equation (\ref{eq:linear_eq}). How can we restrict the solutions to a subset of useful solutions from which we can choose the best one? Here is where the sparse coding model comes into play by offering plausible input images. By putting eq. (\ref{eq:sparse_coding}) in eq. (\ref{eq:linear_eq}) we obtain:
%This is when the sparse coding model comes into play by providing a plausible model for the kind of input images that we are searching for. By replacing equation (\ref{eq:sparse_coding}) in equation (\ref{eq:linear_eq}), we obtain:
\begin{equation}\label{eq:lin_sparse}
    \mathbf{y} = \mathbf{\Phi}\mathbf{D}\mathbf{s} = \tilde{\mathbf{D}}\mathbf{s}, \mbox{ with } \|\mathbf{s}\|_0 \le K \ll J,
\end{equation}
where $\mathbf{D}\in\mathds{R}^{I\times J}$ ($J\ge I$) and $\tilde{\mathbf{D}}=\mathbf{\Phi}\mathbf{D}\in \mathds{R}^{M\times J}$. We then can solve equation (\ref{eq:lin_sparse}) for sparse vectors $\mathbf{s}\in\mathds{R}^J$ by applying any of the available sparse solvers discussed in section \ref{sec:sparse_coding}. Finally, the solution of the original problem can be estimated as $\hat{\mathbf{x}}=\mathbf{D}\mathbf{s}$.

\begin{figure}[!t]
\centering
  \includegraphics[width=0.80\columnwidth]{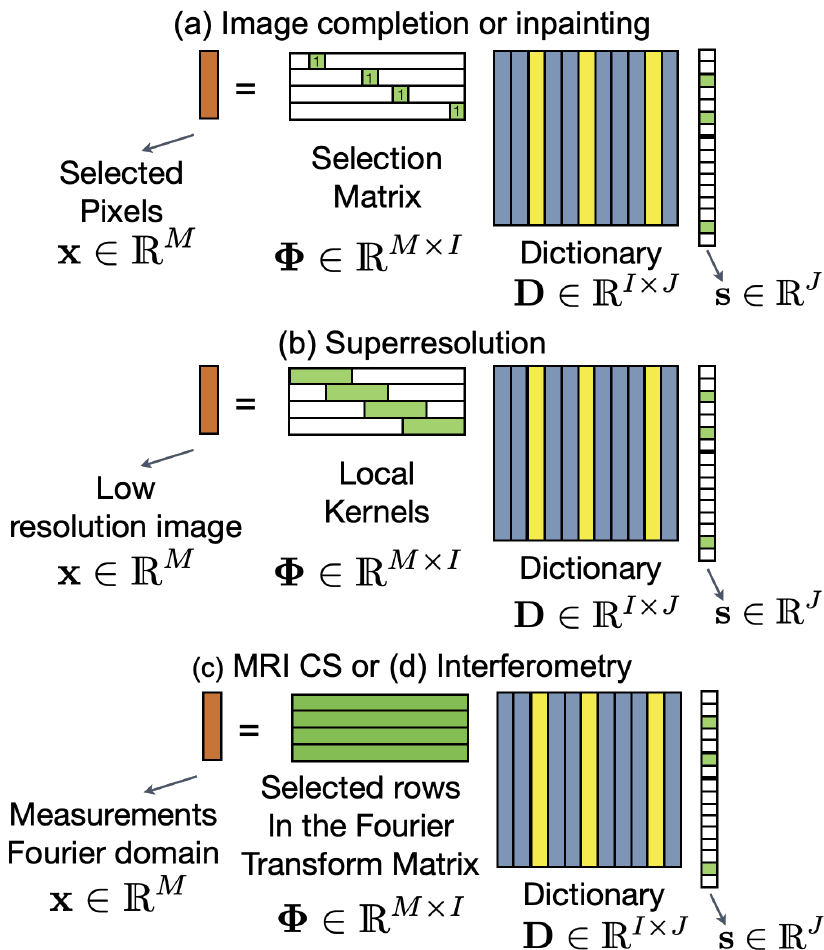}
  \caption{\footnotesize Mathematical formulation of ill-posed linear inverse problems assuming the sparse coding model: (a) inpainting, (b) superresolution, (c) MRI CS  and (d) interferometry.} 
  \label{fig:Fig_3}
\end{figure}

The linear measurement model can be found in several imaging problems (Fig. \ref{fig:Fig_3}): (a) Image completion or inpainting, where only a subset of the pixels are available and the task is to estimate missing pixels \citep{Mairal2009}; (b) Superresolution, when we want to obtain a high-resolution image from its low-resolution version and the measurements are the result of some local averaging operation \citep{Milanfar2010}; (c) Magnetic Resonance Imaging (MRI) CS, where instead of sampling the full space in the Fourier domain, we want to reconstruct the image from incomplete samples in the Fourier domain \citep{Lustig2008}; and (d) Interferometry, a technique used to combine signals from multiple radio telescopes or antenna elements to create a virtual telescope with a much larger aperture than any individual instrument. The mathematical formulation of interferometry is similar to MRI CS in the sense that the goal is to reconstruct an image from its incomplete measurements in the Fourier domain \citep{Wiaux2009}.

\section{The inverse problem in Microwave Tomography}
MicroWave tomography (MWT) is a low-cost and non-invasive imaging technique that uses microwave signals to generate images of the dielectric properties of an object or tissues. 

\begin{figure}[!t]
\centering
  \includegraphics[width=0.8\columnwidth]{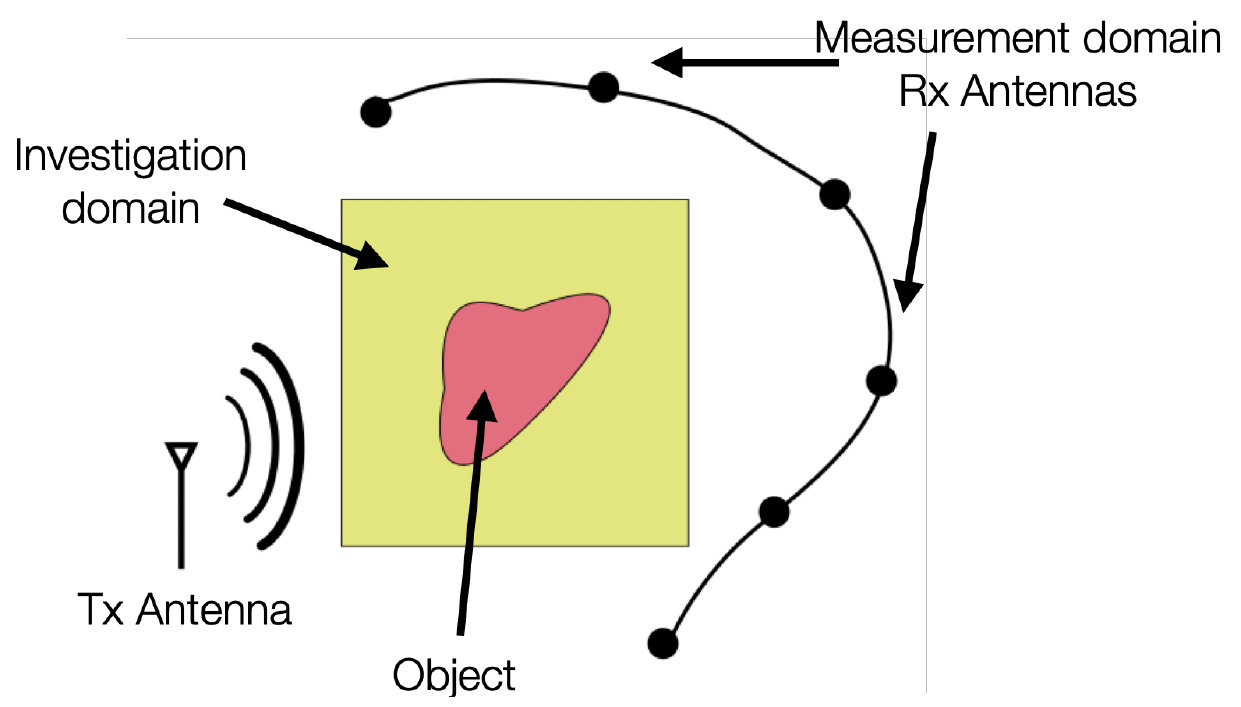}
  \caption{\footnotesize Microwave Tomography (MWT) setting: an object is illuminated with microwave signals while an array of receivers measures the scattered electromagnetic field allowing for a reconstruction of the electromagnetic properties of the object under study.} 
  \label{fig:Fig_4}
\end{figure}

In MWT, a source (transmitter-Tx) emits a microwave signal that pass through the object or body being imaged and is detected by an array of receivers-Rx, which measure the signal magnitude and phase at different points around the object (Fig. \ref{fig:Fig_4}). The collected data is then processed using mathematical algorithms to create a 3D image of the object's internal structure \citep{Pastorino2010}.

MWT has many potential applications, including medical imaging, as it can be used to detect and monitor diseases such as breast or lung cancer. It can also be used in industrial applications, such as monitoring materials in pipelines, metallic silos or detecting material defects in wood industry.

In a two-dimensional (2D) setting, with a homogeneous medium background having permittivity $\epsilon_0$ and permeability $\mu_0$, we are interested in estimating the relative permittivity $\epsilon_r = \epsilon/\epsilon_0$ of a nonmagnetic scatterer (object) located in the domain of interest $\mathcal{D}\in \mathds{R}^2$. We consider $N_{inc}$ transmitters and $N_{rec}$ receivers. We applied Maxwell equations to obtain the Electric-field integral equation \citep{Chen2018}, which after discretization of the domain into $I = M\times M$ pixels through the Methods of Moments (MoM) \citep{Newman1991}, is converted to the following matrix equations:
%After discretizing the domain into $I = M\times M$ pixels and by applying Maxwell equations ({\color{red} Electric-field integral equation EFIE?}) through the Methods of Moments (MoM) \citep{Newman1991} we obtain the following matrix equations of the scattering problem \citep{Chen2018}:
\begin{eqnarray}
 \mathbf{E}^t &=& \mathbf{E}^i + \mathbf{G}_D\mathbf{\Lambda}\mathbf{E}^t, \label{eq:Lip-Sch}\\
\mathbf{E}^s &=& \mathbf{G}_S\mathbf{\Lambda}\mathbf{E}^t, \label{eq:measurements}
\end{eqnarray}
where $\mathbf{E}^t \in \mathds{C}^{I \times N_{inc}}$ contains the {\it Total Electric Field}; $\mathbf{E}^i \in \mathds{C}^{I \times N_{inc}}$ contains the {\it Incident Electric Field}; $\mathbf{E}^s \in \mathds{C}^{N_{rec} \times N_{inc}}$ contains the {\it Measured Electric Field} at the receivers; $\mathbf{G}_D \in \mathds{C}^{I\times I}$ and $\mathbf{G}_S \in \mathds{C}^{N_{rec}\times I}$ are the matrix versions of the corresponding 2D free space Green's function; and $\mathbf{\Lambda} \in \mathds{C}^{I\times I}$ is a diagonal matrix whose main diagonal entries contain the information of the relative permittivity ${\epsilon_r}_i$ at $i$-th pixel as follows:
\begin{equation}
\mathbf{\Lambda}_{i,i} = \lambda_i = -j\Omega ({\epsilon_r}_i - 1)\epsilon_0 \Delta_a,
\end{equation}
with $\Omega = \omega\sqrt{\mu_0\epsilon_0}$, $\omega$ being the angular frequency and $\Delta_a$ is the area of each pixel.

To reconstruct the complex relative permittivity map ${\epsilon_r}_i = \epsilon_i / \epsilon_0$, for $i=1,2,\dots, I$,
%($\epsilon_i = \epsilon_0 {\epsilon_r}_i$ for $i=1,2,\dots, I$) 
given the measurements of the electric field $\mathbf{E}^s$ involves solving the inverse problem of finding the diagonal matrix $\mathbf{\Lambda}$ such that equations (\ref{eq:Lip-Sch}) and (\ref{eq:measurements}) are satisfied. It is well known that this is a highly nonlinear and unstable inverse problem \citep{Chen2018, Pastorino2010}.

\subsection{Classical approaches}
In general, solving the equations (\ref{eq:Lip-Sch}) and (\ref{eq:measurements}) for $\mathbf{\Lambda}$ has not an analytical form and a numerical approach is needed. Classical methods are classified into non-iterative and iterative algorithms.

Non-iterative methods assume some approximation of the equations, for example: (1) the object being imaged scatters the microwave radiation in a linear manner (Born approximation); (2) a simple but higher-order approximation than the Born method is used (Rytov and extended Born approximations); (3) the induced current is proportional to the back-propagated field which is computed through the adjoint operator of the Green's function (Back-Propagation (BP)) \citep{Chen2018}. 

On the other hand, iterative inversion methods are usually based on ``exact'' models and iteratively refine the solution until the difference between the measured scattered field and the predicted scattered field is minimized. Examples of iterative methods are: the Distorted Born Iterative Method (DBIM), the Contrast Source Inversion (CSI) method, the Contrast Source Extended Born (CS-EB) method, the Subspace-Based Optimization Method (SOM), and others \citep{Chen2018}.

\subsection{Machine Learning based methods}
Classical MWT inverse methods suffers from low spatial resolution, making this technology not yet suitable for some applications such as tumor detection. Recently, new inversion methods were explored by applying deep neural networks trained on large collections of pairs of data samples $\{\mathbf{E}^s, \mathbf{\Lambda} \}_t$ for $t=1,2,\dots,T$ so that a neural network learns how to map measurements $\mathbf{E^s}$ to permittivity map images \citep{Wei2019,Li2019}. Then, when a new measurement is presented to the neural network, it gives an approximated solution to the MWT inverse problem. While the results seemed to be promising, exhibiting better spatial resolution than classical methods, NNs are used as black boxes which has several drawbacks: they are not reliable, there is not an interpretation of the results \citep{Weld2018} and the measurements physical model is not used explicitly.

\subsection{A sparse coding approach}
We propose to learn a sparse coding model of the contrast maps $\mathbf{x} = (\bm{\epsilon}_r - \mathbf{1}) \in \mathds{R}^I$, where $\bm{\epsilon}_r \in \mathds{R}^I$ is the vector of relative permittivities, such that, by applying CS theory, we can restrict the subset of admitted solutions and solve the inverse problem.

To solve the inverse scattering problem, we need to find a diagonal matrix $\mathbf{\Lambda}$ and a total field matrix $\mathbf{E}^t$ such that equations (\ref{eq:Lip-Sch}) and (\ref{eq:measurements}) are satisfied. Here, we propose to iteratively refine these variables such that the following cost function is minimized:
\begin{gather}\scriptsize \label{eq:cost}
C(\mathbf{\Lambda}, \mathbf{E}^t) =  
\alpha^2 \|(\mathbf{I} - \mathbf{G}_D\mathbf{\Lambda})\mathbf{E}^t - \mathbf{E}^i\|^2_F + \nonumber \\
(1-\alpha^2) \| \mathbf{G}_S\mathbf{\Lambda}\mathbf{E}^t - \mathbf{E}^s \|^2_F,
\end{gather}
%\begin{equation}\scriptsize \label{eq:cost}
%C(\mathbf{\Lambda}, \mathbf{E}^t) = 
%\alpha^2 \|(\mathbf{I} - %\mathbf{G}_D\mathbf{\Lambda})\mathbf{E}^t - \mathbf{E}^i\|^2_F +
%(1-\alpha^2) \| \mathbf{G}_S\mathbf{\Lambda}\mathbf{E}^t - \mathbf{E}^s \|^2_F,
%\end{equation}
where $\alpha$ is a hyper-parameter that can be tuned by cross-validation. The idea is to minimize $C(\mathbf{\Lambda}, \mathbf{E}^t)$ keeping, at the same time, a sparse representation of the permittivity map using a dictionary $\mathbf{D}\in \mathds{R}^{N\times J}$ trained previously on a large collection of images.

We minimize $C(\mathbf{\Lambda}, \mathbf{E}^t)$ by alternately optimizing it for one of the two variables keeping the other fixed, arriving in both cases at classical least squares problems. However, optimizing for $\bm{\Lambda}$ with fixed $\mathbf{E}^t$ requires to impose the sparse coding constraint on $\bm{\Lambda}$, which can be done by using any available sparse coding algorithm as described in section \ref{sec:sparse_coding}.

%\subsection{Simulation results}

Here, we present experimental results on synthetic data (ground truth) generated by creating circles with a constant permittivity within a 2D grid of size $32\times 32$, i.e. $I=32^2=1,024$. The circles were randomly placed and varied in size as in \citep{Wei2019}. 
%using ground truth permittivity maps using circles with constant permittivity inside and having random sizes and locations within a 2D grid of size $32\times 32$, i.e. $I=32^2=1,024$ \citep{Wei2019}. 
We trained a dictionary matrix $\mathbf{D}\in \mathds{R}^{1,024 \times 4,096}$ using the learning algorithm developed in \citep{Mairal2009} on a dataset with $T=250,000$ images. We consider $N_{inc}=16$ transmitter positions, $N_{rec}=32$ receivers, frequency $f=400$ Mhz, $\epsilon_0=8,85\times 10^{-12}$ farad/m and hyper-parameter $\alpha=0.1$. To compute the sparse coefficients we applied the Fast Iterative Shrinkage-Thresholding Algorithm (FISTA) \citep{Beck2009}, which employs $\ell_1$ regularization.

\begin{figure}[!t]
\centering
  \includegraphics[width=1.0\columnwidth]{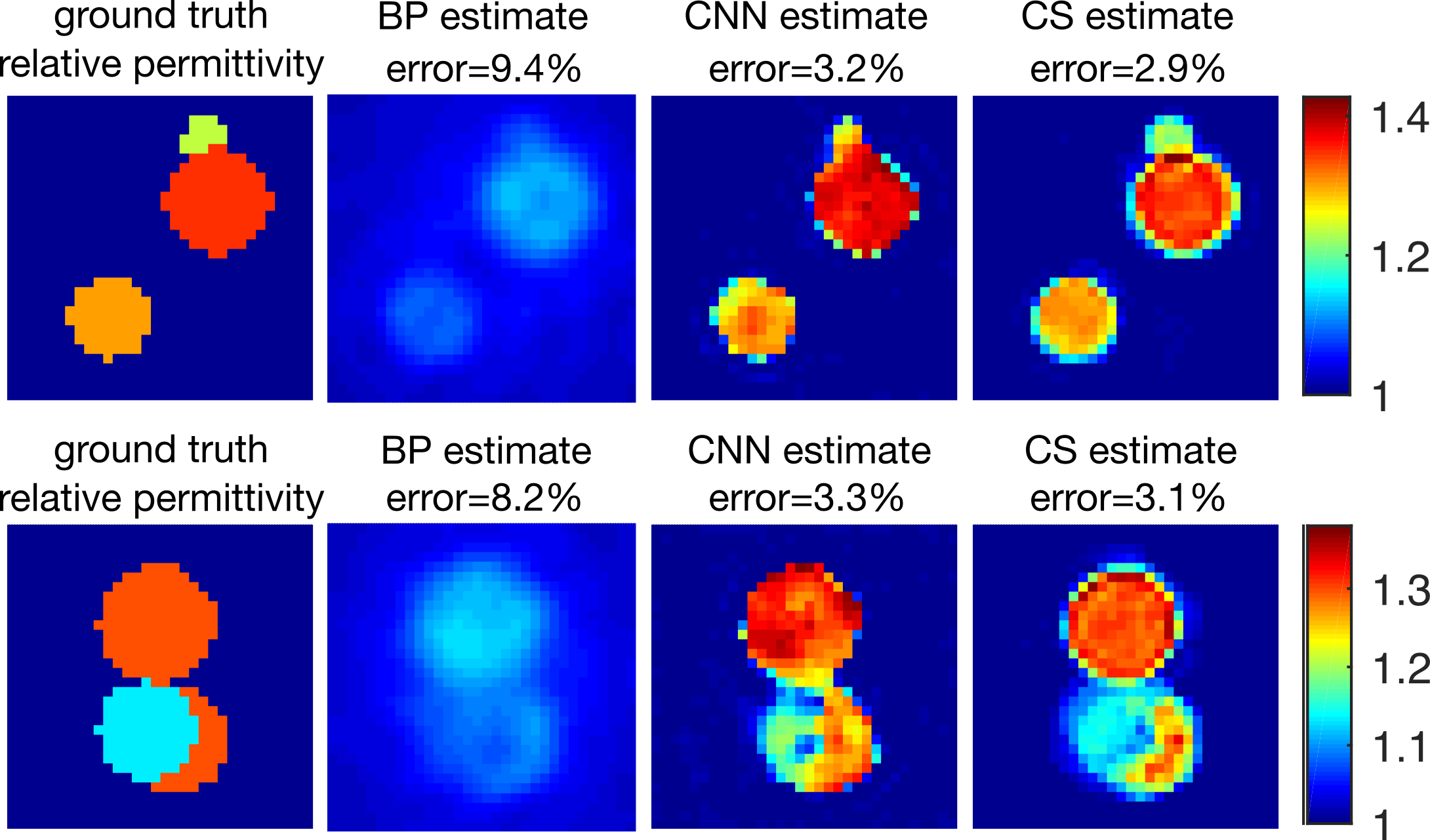}
  \caption{\footnotesize Two examples comparing the results of our method (CS) with a classical direct algorithm (BP) and a state-of-the-art method based on neural networks (CNN).} 
  \label{fig:Fig_5}
\end{figure}

For comparison purposes we also replicated results from \citep{Wei2019} and trained a U-Net neural network to estimate unseen input images based on their measurements. In Fig. \ref{fig:Fig_5}, two examples of relative permittivity map reconstruction of unseen images are shown comparing the results of our method (CS) with a classical direct method (BP) \citep{Chen2018} and the state-of-the-art method based on neural networks (CNN) \citep{Wei2019}. The CS method provided the lowest error and also visually more accurate reconstructions.

% To include previous results from Irastorza group. 

\section{Conclusions}
The theory of CS developed in recent years demonstrated that the sparse coding model provides an effective and practical way to solve ill-posed inverse problems in the linear setting. Moreover, today we have algorithms and theory supporting them in terms of achievable accuracy and computation complexity. However, the application of CS theory to non-linear inverse problems, such as the case of MWT imaging, was not fully explored in the past \citep{Olivieri2017}. In this work, we showed that we can convert the initial non-linear problem into two linear sub-problems by using an Alternate Least Squares (ALS) approach and apply CS methods. Our preliminary results showed that better reconstructions can be obtained with this approach compared with classical methods such as classical BP and recent deep-learning methods. 

{\small
\textbf{\center Acknowledgments:} This work was supported by grants PICT 2020-SERIEA-00457 and PIP 112202101 00284CO (Argentina).}

% Can sparse coding help to improve image reconstructions in Microwave Tomography?
% What are the similarities and differences of solving the inverse linear problem compared with the microwave tomography inverse problem?
% Future directions of research

%\begin{figure}[!t]
%  \includegraphics[width=\columnwidth]{example-fig}
%  \caption{Example of a simple single-column figure. Don't put this
%    too early in the document since we don't want it to go in the
%    first column.}
%  \label{fig:simple}
%\end{figure}

\end{document}